\documentclass[review]{elsarticle}
\usepackage{amssymb}
\usepackage{lineno,hyperref}
\modulolinenumbers[5]
\journal{Nuclear Science and Techniques}

\begin{document}
\begin{frontmatter}
\title{A segmented conical electric lens for optimization of the beam spot of the low-energy muon facility at PSI:
a Geant4 simulation analysis}

\author[mymainaddress,mysecondaryaddress,mythirdaddress]{Ran Xiao\corref{mycorrespondingauthor}}
\cortext[mycorrespondingauthor]{Corresponding author}
\ead{xran@mail.ustc.edu.cn}

\author[mythirdaddress]{Elvezio Morenzoni}

\author[mythirdaddress]{Zaher Salman}

\author[mymainaddress,mysecondaryaddress]{Bangjiao Ye}

\author[mythirdaddress]{Thomas Prokscha\corref{mycorrespondingauthor}}
\cortext[mycorrespondingauthor]{Corresponding author}
\ead{thomas.prokscha@psi.ch}

\address[mymainaddress]{State Key Laboratory of Particle Detection and Electronics, University of Science and Technology of China, Hefei 230026, P. R. China}
\address[mysecondaryaddress]{Department of Modern Physics, University of Science and Technology of China, Hefei 230026, P. R. China}
\address[mythirdaddress]{Laboratory for Muon Spin Spectroscopy, Paul Scherrer Institut, CH-5232 Villigen PSI, Switzerland}

\begin{abstract}
The low energy muon (LEM) facility at PSI provides nearly fully polarized positive muons with tunable energies in 
the keV range to carry out muon spin rotation (LE-$\mu$SR) experiments with nanometer depth resolution
on thin films, heterostructures and near-surface regions. The low energy muon beam is focused and transported to
the sample by
electrostatic lenses. In order to achieve a minimum beam spot size at the sample position, and to enable the
steering of the beam in horizontal and vertical direction, a special electrostatic device has been implemented
close to the sample position. It consists of a cylinder at ground potential, followed by four conically shaped
electrodes which can be operated at different electric potential. In LE-$\mu$SR experiments an electric field
at the sample along the beam direction can be applied to accelerate/decelerate muons to different energies 
(0.5-30 keV). Additionally, a horizontal or vertical magnetic field can be superimposed for transverse or
longitudinal field $\mu$SR experiments. The focusing properties of the conical lens in the presence of these
additional electric and magnetic fields have been investigated and optimized by Geant4 simulations. Some
experimental tests were also performed which show that the simulation well describes the experimental setup.
\end{abstract}


\end{frontmatter}

\section{\label{sec1}Introduction}
The muon spin rotation/relaxation/resonance ($\mu$SR) technique is a versatile local probe technique to investigate the 
physical properties of superconductors, magnetic systems, semiconductors and organic materials \cite{Reffer1}.
Polarized muon beams for $\mu$SR applications are usually produced at medium energy (0.5 - 3 GeV) proton
accelerators. These muons have kinetic energies of the order of MeV and penetrate deeply into a
sample (mm to cm). Therefore, $\mu$SR experiments using these MeV muons can only study bulk materials. 
To overcome these limitations and to extend $\mu$SR to the investigation of thin films, 
PSI developed and operates the low energy muons beam facility (LEM) where a cryogenic moderation 
method \cite{Reffer2,Reffer3,harshman1987,Reffer11,Reffer10} is used to generate nearly fully polarized
positive muons with tunable energies in the range of eV to several keV. 
Up to now the LEM facility at PSI has played a leading role in low energy muon experiments, 
extending the $\mu$SR technique to the investigation of nano-materials, layer interfaces, thin films 
and near-surface regions
\cite{Reffer4,Reffer5,Reffer6,Reffer7,Reffer8,Reffer9,boris2011,hofmann2012,stilp2014,saadaoui2015,al2015,anghinolfi2015,flokstra2015}.

The LEM facility is located at the $\mu$E4 beam line, which is a hybrid-type large acceptance channel to generate an 
intense beam of so-called surface muons (positive muons, $\mu^+$, originating from pions decaying at rest close 
to the surface of the pion/muon target, with a kinetic energy of $\sim$ 4 MeV) \cite{Reffer10}.  
The intensity of the surface muon beam at the exit of the $\mu$E4 beam line is about 4.6$\times$10$^8$/s at a
proton current of 2.2 mA. This represents at the moment the highest continuous surface muon flux in the world. 
About 40$\%$ of the beam is focused onto the cryogenic moderator target.  
Using a wide-band-gap van der Waals solid gas (s-N$_2$, s-Ar) a moderation efficiency 
($\frac{N^{out}_{eV}}{N^{in}_{MeV}}$ ) between 10$^{-5}$ and 10$^{-4}$ is achieved \cite{Reffer2,harshman1987,Reffer5}.  

The moderation of the muons from about 4 MeV to 10 eV is achieved within 10 ps, such that their initial high
polarization is conserved \cite{Reffer11}. The moderator consists of a 200-300 nm thick Ar layer (capped by a 
10 nm thin N$_2$ layer) deposited on a thin Ag foil ($\sim$125 $\mu$m) \cite{Reffer12} which is held by a cryostat
at a temperature below 20 K. The mean energy of the moderated muons is about 15 eV with a width of about 20 eV 
(full width at half maximum, FWHM) \cite{Reffer13}. These moderated muons can be re-accelerated by applying a high 
positive potential of up to 20 kV to the moderator \cite{Reffer14}. After acceleration they are transported by 
electro- and magneto-static beam elements to the sample position. The rate of moderated muons at the sample is 
up to 4.5$\times$10$^3$/s. 

By tuning the high voltage of the moderator and the acceleration/deceleration high voltage
at the sample, low-energy positive muons (LE-$\mu^+$) with tunable implantation energies between 
0.5 and 30 keV are obtained, corresponding to mean implantation depths ranging from a few nm to a few
hundred nm in solid materials \cite{Reffer15}. In addition to the electric acceleration/deceleration field
at the sample an external magnetic field -- either parallel or perpendicular to the muon momentum -- can be
applied for transverse and longitudinal field $\mu$SR measurements. These fields may influence the beam
spot size and position. A special optical element with four conically shaped segments (also called ring anode, RA) 
is used to focus the beam onto the sample. Its focusing and steering effects in combination with the
applied electric and magnetic fields have been investigated using the musrSim simulation 
package \cite{Reffer18} which is based on Geant4 \cite{Reffer19,Reffer20}. 
Using these simulations we optimize the settings of RA for the various magnetic fields and varying 
implantation energies (electric fields), and compare them with experimental data. 
These parameters allow to run the experiment with optimized beam
transport onto the sample under different magnetic and electric field configurations.
The results of this analysis are presented in this paper.
\begin{figure}[t]
\includegraphics[width=8.5cm]{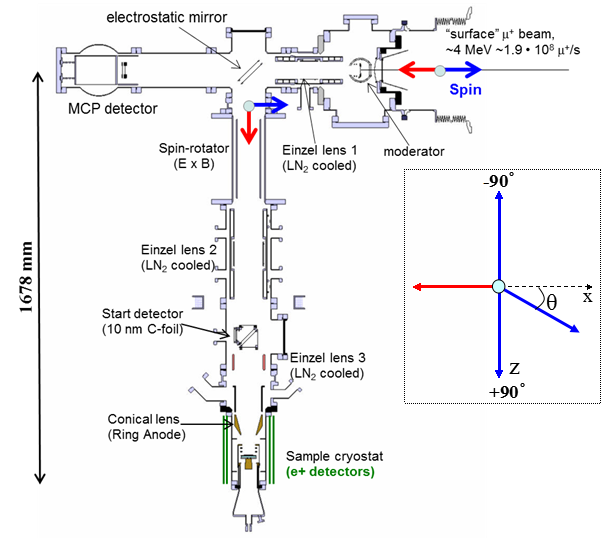}
\centering
\caption{The layout of the low energy muon beam line for LE-$\mu$SR   experiments at PSI.  The red and blue arrows in the figure represent the muon's momentum and direction of spin polarization, respectively. 
$\theta_{spin}$ represents the spin angle with  respect to the initial direction of the 4-MeV muon beam, corresponding to $-x$ direction in the simulation.}
\label{FIG1}
\end{figure}
\section{\label{sec2} Analysis of the effect of the external magnetic and electric fields at the sample position}

\section{\label{sec.II} Setup of the low energy muon beam}
A schematic of the LEM apparatus is shown in Fig.~\ref{FIG1}. 
The low-energy muon beam is extracted from the cryogenic moderator with a quadratic area of $30\times30$ mm$^2$. 
Adjustable positive high voltages applied to the moderator and to a set of grids generate an accelerating electric 
field to re-accelerate the moderated muons from eV to higher energies, typically between 10 keV and 20 keV 
(low energy muons). Only a small fraction of surface muons is moderated to eV energies, and most of the surface
muons stop in the moderator target or leave the moderator with keV to hundreds of keV energies (fast muons). 
An electrostatic mirror with an angle of 45$^\circ$ with respect to the muon momentum is used to separate low energy 
from fast muons. After being focused by einzel lens 1 (L1) the low energy muons are deflected by 90$^\circ$ with 
respect to the initial muon direction, while the fast ones continue in the direction of the MCP detector. 
After the deflection the spin polarization is perpendicular to the muon momentum.
\begin{figure}[ht]
\includegraphics[width=6cm]{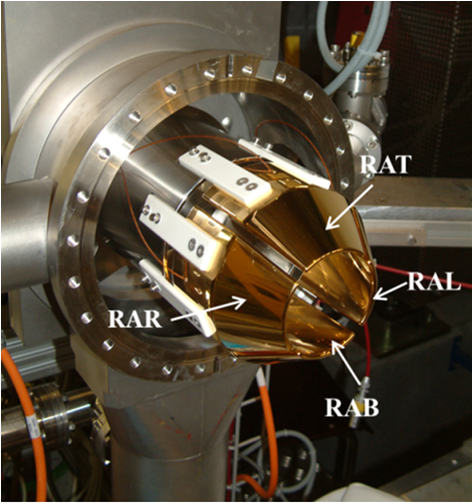}
\centering
\caption{The four segments of the ring anode RA: the top segment RAT, the bottom segment RAB, the right segment RAR, 
and the left segment RAL, attached to a grounded stainless steel cylinder by pairs of macor insultators. The 
left and right segments of the RA sectors are named with respect to the muon beam direction.}
\label{FIG2}
\end{figure}
A spin rotator with crossed static magnetic and electric fields $\vec{E} \times \vec{B}$ can be used to rotate the
muon spin parallel or anti-parallel to its momentum \cite{Reffer16}. This enables to carry out longitudinal field
measurements (LF-$\mu$SR, where the muon spin is parallel or anti-parallel to the applied field at the sample). 
The LE-$\mu^+$ spin angle ($\theta_{spin}$) can be changed between -90$^\circ$ to +90$^\circ$ with
respect to the initial direction of the 4-MeV muon beam by tuning the magnetic and electric fields in the spin rotator, 
such that the ratio E/B matches the velocity of the muons.
Downstream of the spin rotator the einzel lens 2 (L2) focuses the muon beam onto the retractable start detector. 
This detector provides the fast start timing signal which is necessary for $\mu$SR experiments at a continuous muon beam.
Its time resolution is about 1~ns, the detection efficiency for keV muons is $\sim 80$\%, and it introduces an increase
of transverse phase space due to multiple scattering, and an increase of the initial width of the energy distribution 
of 20~eV FWHM to about 1~keV FWHM. The detector is described in detail in Ref.~\cite{morenzoni1997,khaw2015geant4}.
The final beam size at the sample position is determined by lens 3 (L3) and the strongly focusing conical lens RA. 
It consists of the four segments RAT, RAB, RAL and RAR, see Fig.~\ref{FIG2}.
These segments are made of gold-plated, polished copper and are attached to a grounded stainless steel cylinder by
pairs of macor or sapphire insulators. In contrast to an einzel lens the conical lens can be placed closer
to the sample position, thus allowing for stronger focusing and a smaller beam spot. The beam can be shifted 
in horizontal and vertical direction by applying potential differences between the segments.

Compared to other low-energy particle beams with millimeter size beam extension and low emittances, the PSI
low-energy muon beam has a large phase space due to the large source size of $30\times30$ mm$^2$, the
initial  $\cos\theta$ angular distribution (see Sec.~\ref{sec.III}), and the additional increase of phase space 
after passing the beam through the 10-nm-thin carbon foil of the start detector. This makes beam transport with
a final small beam spot much more challenging.

\section{\label{sec.III} Optimization of RA focusing with applied external magnetic and electric fields at the sample}

\begin{figure*}[t]
\includegraphics[width=7.0 cm]{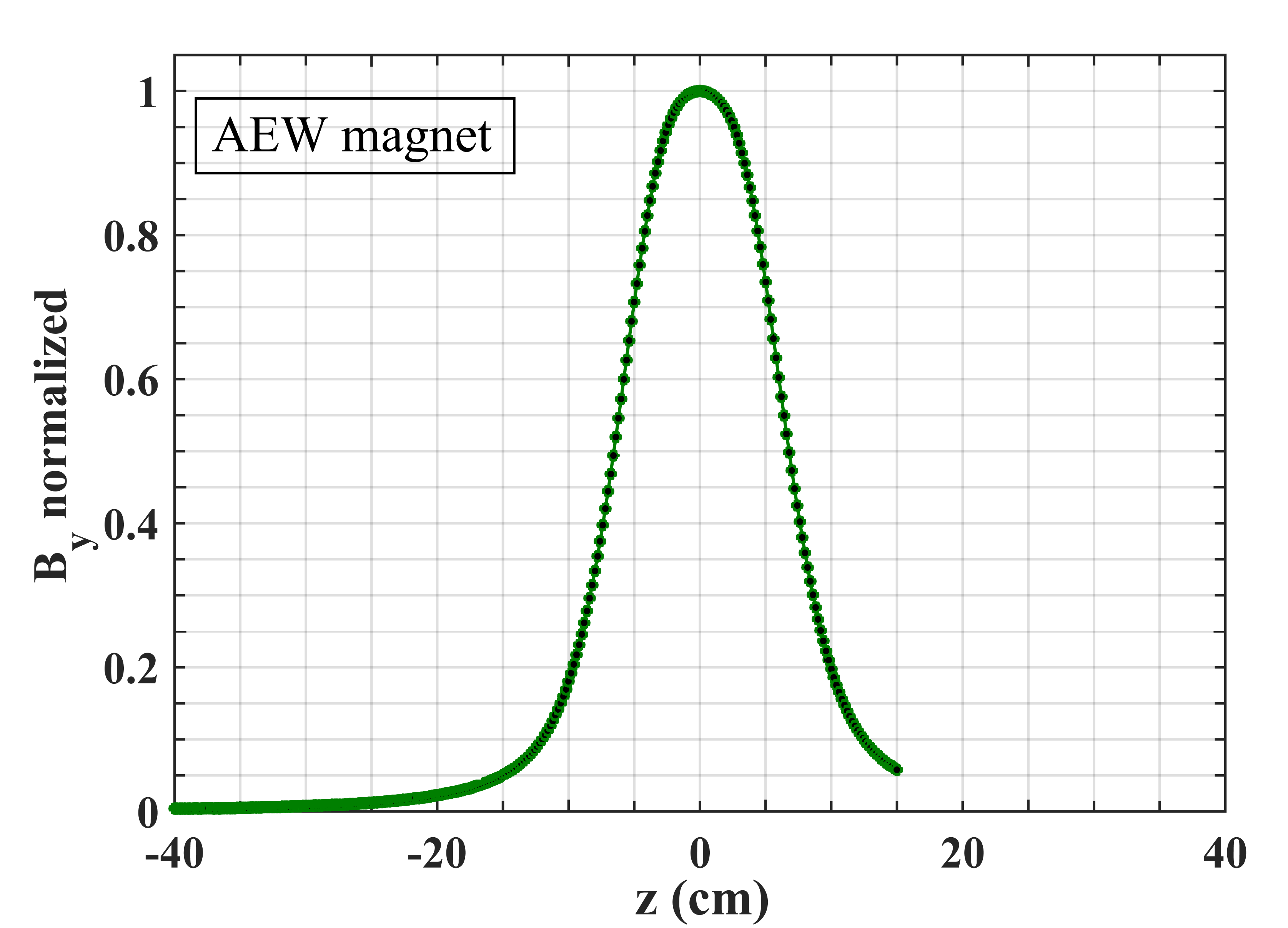}
\hspace{1cm}
\includegraphics[width=7.0 cm]{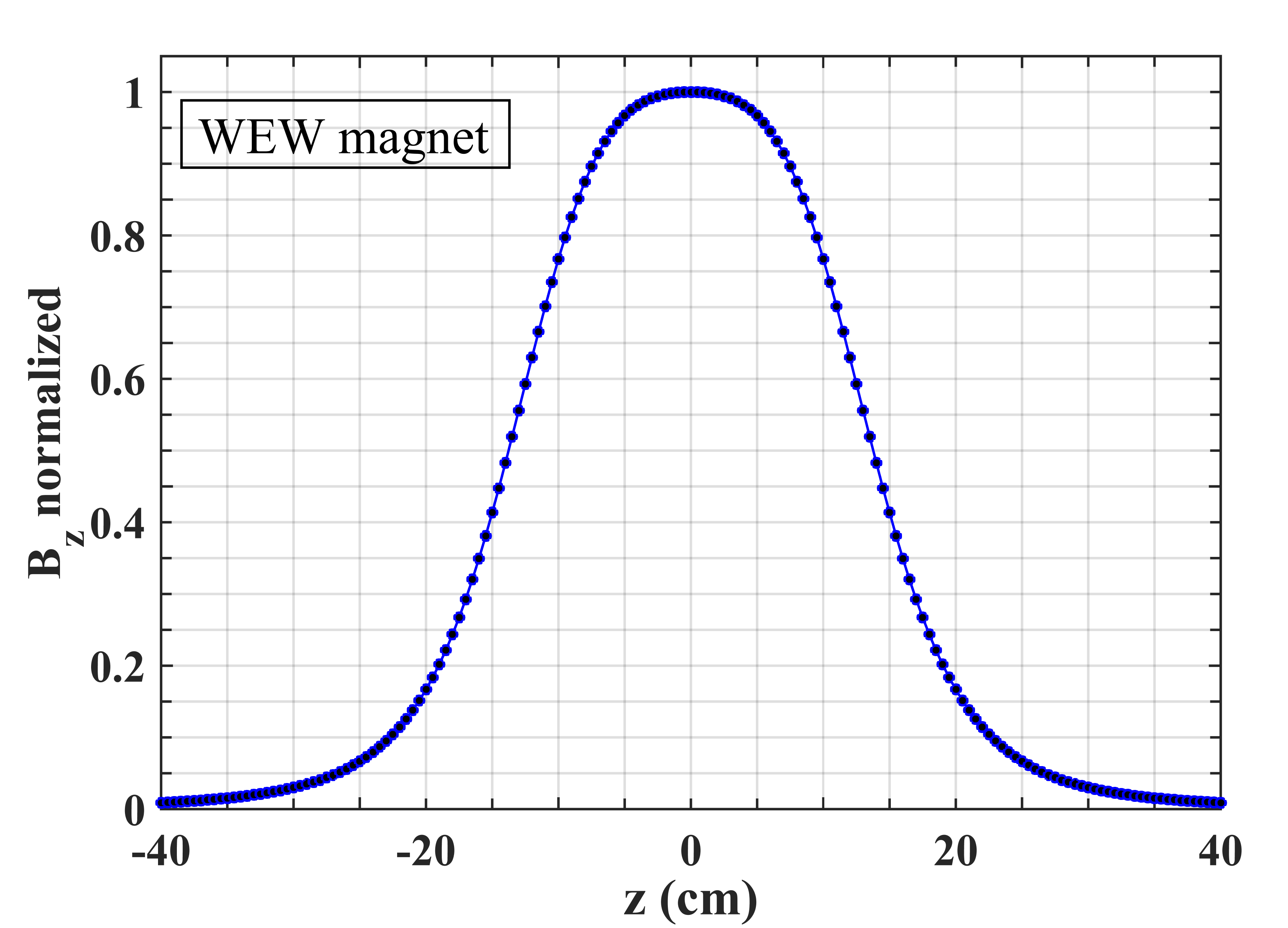}
\centering
\caption{Normalized magnetic field along beam axis z of the AEW magnet (measured) and WEW magnet (calculated).
z = 0 corresponds to the sample position.}
\label{FIG3NEW}
\end{figure*}
\begin{figure}[t]
\centering
\includegraphics[width=7cm]{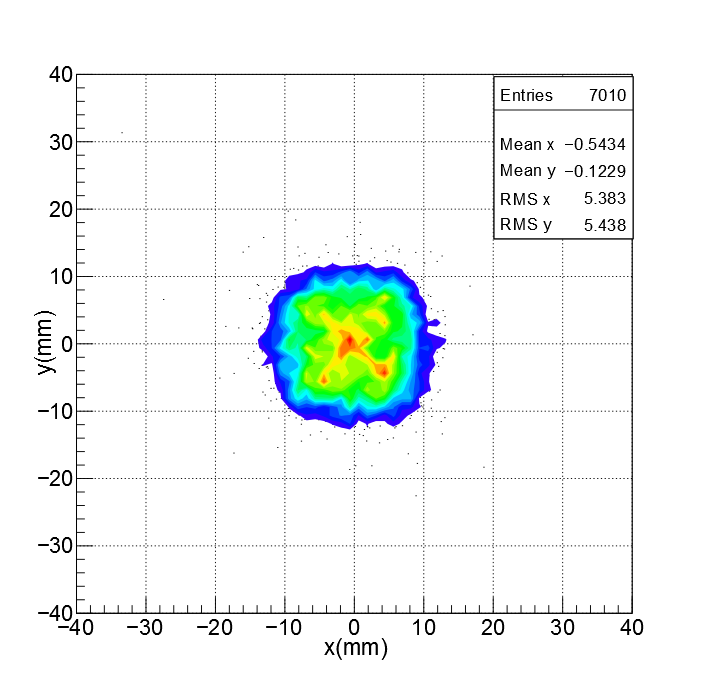}
\centering
\caption{The simulated beam spot without external fields at the sample position using optimized experimental
high voltage settings for the beam line elements at V$_{mod}$=15 kV, $\theta_{spin}$=-10$^\circ$:
  L1=9.0 kV, L2=10.5 kV, L3=11.5 kV and RA=11.9 kV. The initial number of muons at the moderator ($N^{in}$) is $10^4$.}
\label{FIG4}
\end{figure}
\begin{figure*}[t]
\includegraphics[width=14 cm, height=8 cm]{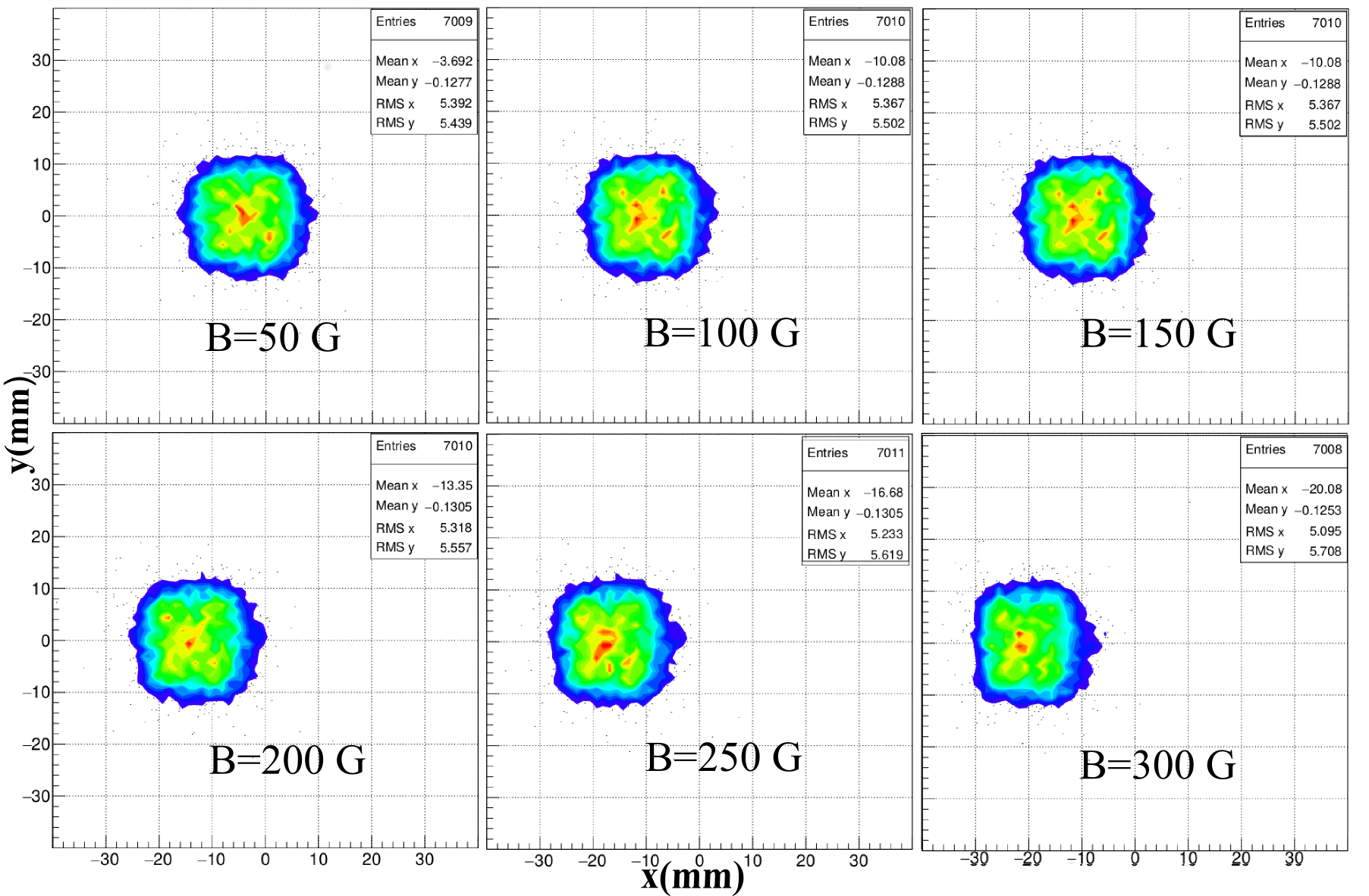}
\centering
\caption{Simulated beam spots at the sample viewed from downstream in different AEW magnetic fields (50 to 300 Gauss) with the same beam element settings as in 
Fig. \ref{FIG2}.}
\label{FIG5}
\end{figure*}
\begin{figure}[ht]
\centering
\includegraphics[width=7 cm]{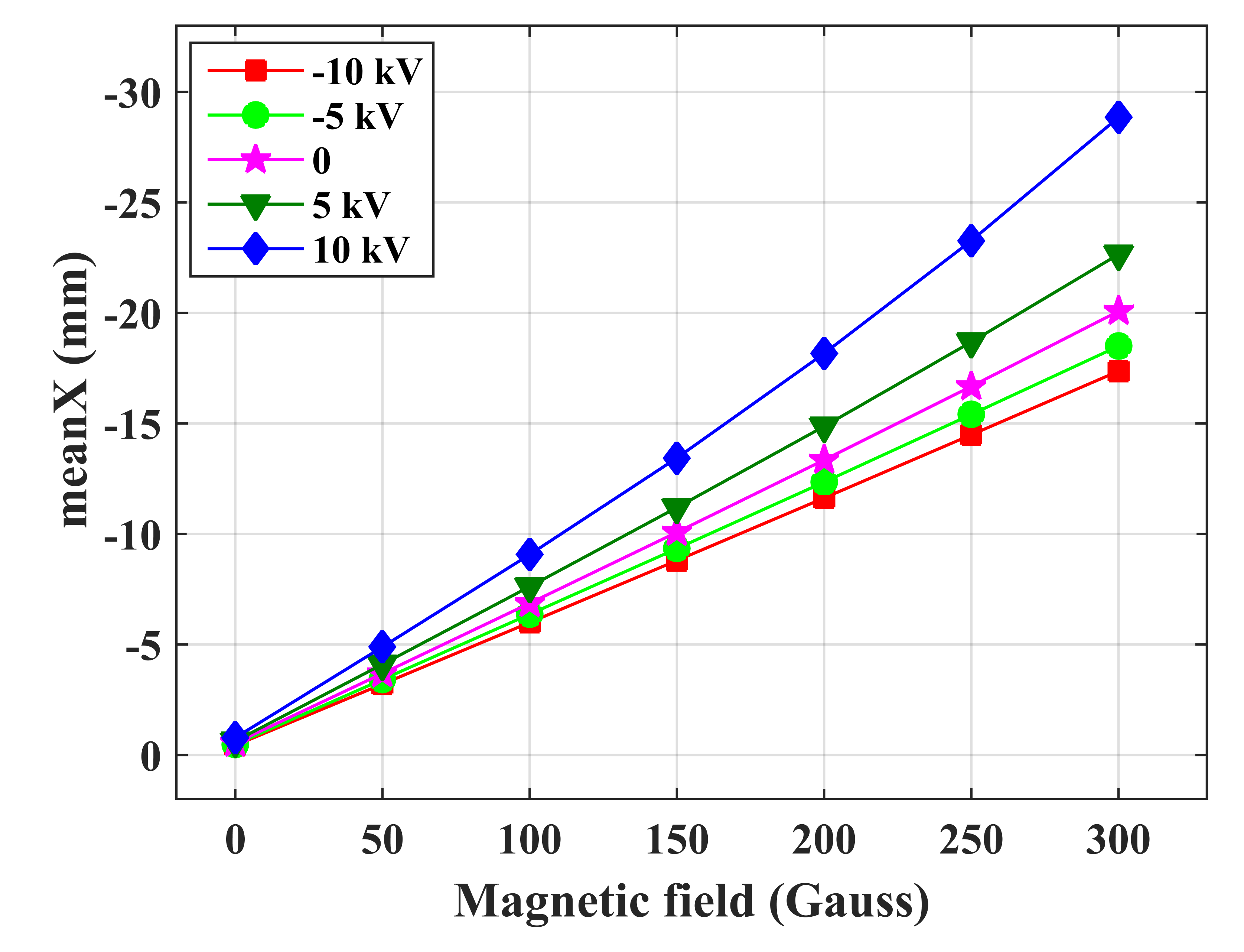}
\centering
\caption{meanX values with electric potentials at the sample from -10 kV to 10 kV 
and AEW magnetic fields from 0 to 300 Gauss (V$_{mod}$ = 15 kV)}
\label{FIG6}
\end{figure}
In the LEM beam Geant4 simulation, the initial muon beam is started at the moderator, at the edge of the
high-voltage acceleration region. The moderator is located at a distance of 499 mm upstream to the center
of the electrostatic mirror. We assume homogeneous electric fields in the acceleration volume. In the simulation
we studied the beam transport for various beam energies, i.e. moderator potentials $V_{mod}$: 10~kV, 12~kV, and
15~kV. The initial beam is homogeneously distributed on the moderator area of 30$\times$30 mm$^2$, 
and the initial mean kinetic energy is 15 eV with a FWHM of 20 eV. Due to the homogeneous angular distribution of
slow muons inside the moderator \cite{Reffer12} a $\cos\theta$ angular distribution is generated in the simulation
for muons escaping from the moderator layer, where $\theta$ is the angle of the muon momentum with respect to the
initial beam direction (the $-x$ direction in the simulation). 

Field maps of the beam elements (Spin Rotator, L1, L2, L3 and RA) have been either calculated or
measured \cite{Reffer16,Reffer17,khaw2015geant4}, and are scaled according to the experimental settings.  
In the electrostatic mirror a homogeneous electric field is assumed.

A vertical magnetic field up to 300 Gauss, parallel to the sample surface, is supplied by the AEW magnet for
transverse field $\mu$SR measurements (TF-$\mu$SR, muon spin transverse to the applied field), 
and a magnetic field of up to 3400 Gauss can be applied perpendicular to the sample surface and along the
muon beam direction (WEW magnet, LF- or TF-$\mu$SR, depending on the initial muons spin polarization). The AEW
magnet consists of air-cooled coils (8~A maximum current) wound around a soft-iron yoke with a magnet gap of 154~mm.
Because of this large gap, soft-iron pieces inside the vacuum tube are used to reduce the effective length of the
magnet and its fringe field region, and to increase the maximum available magnetic field at the sample position.
The WEW magnet consists of two water-cooled coils, made of square copper-hollow conductors (600~A maximum current),
in Helmholtz geometry, surrounded
by a soft-iron housing to reduce the stray fields outside of the magnet, and to maximize the available
field at the sample position. The iron housing of the magnet has a length of 285~mm in beam direction, 
and a front face size of 540$\times$540~mm$^2$. The
magnetic field along the beam axis is shown for both magnets in Fig.~\ref{FIG3NEW}. The effective lengths of the
magnets are 152~mm for the AEW magnet, and 295~mm for the WEW magnet.

An acceleration/deceleration high voltage of up to $\pm$12.5 kV can be applied at the sample plate in order
to tune the final implantation energy of the LE-$\mu^+$. 
\begin{figure*}[ht]
\centering
\includegraphics[width=4.5cm]{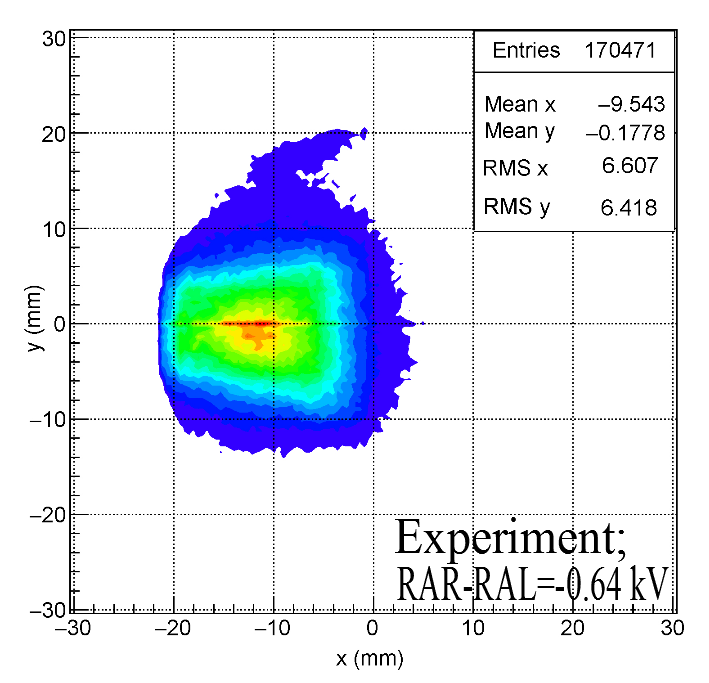}
\includegraphics[width=4.5cm]{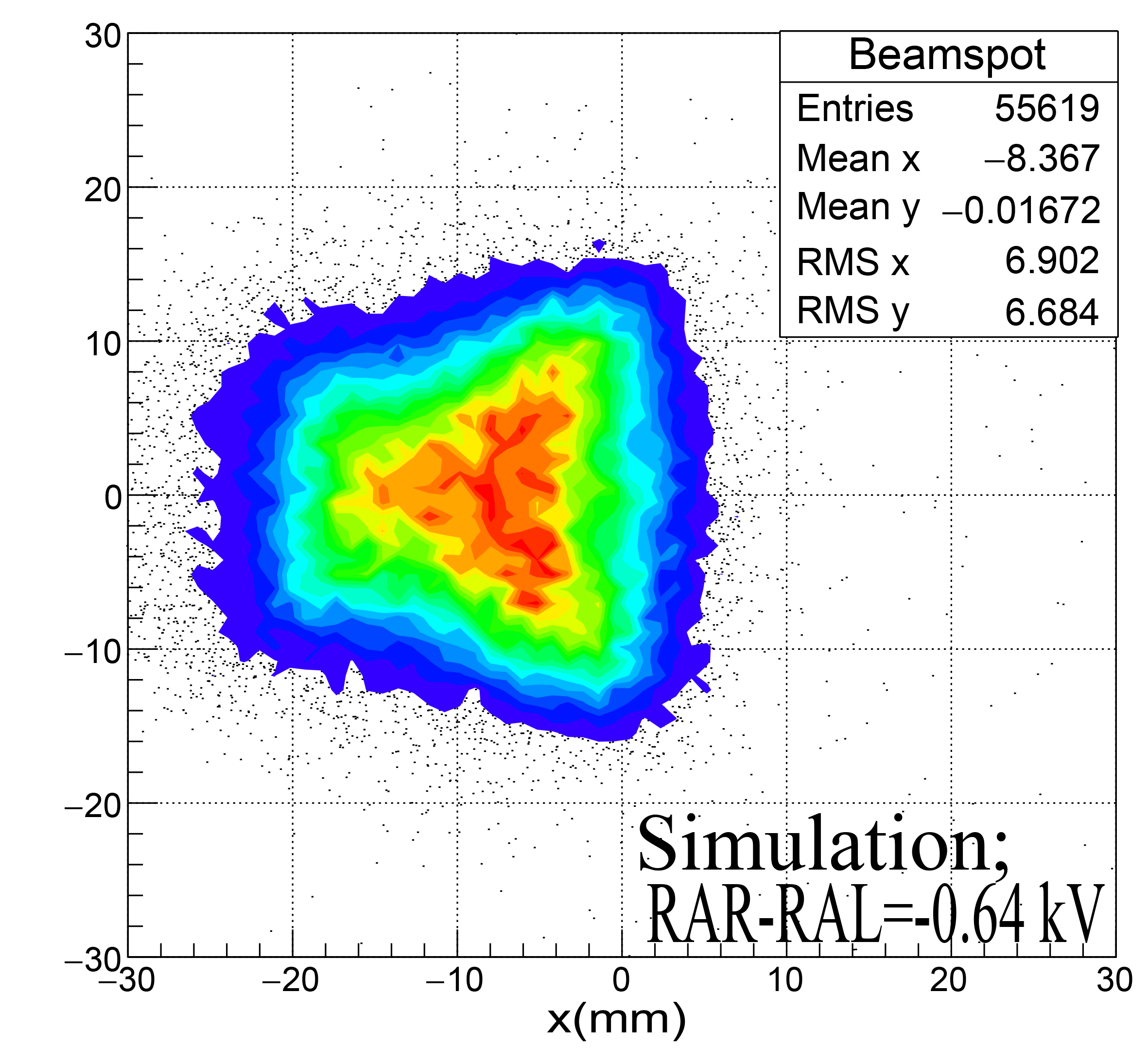}\\
\includegraphics[width=4.5cm]{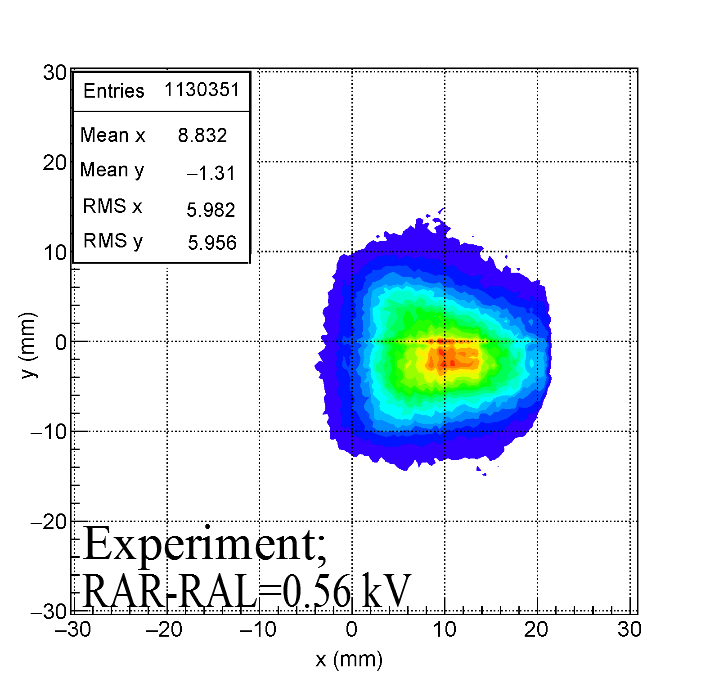}
\includegraphics[width=4.5cm]{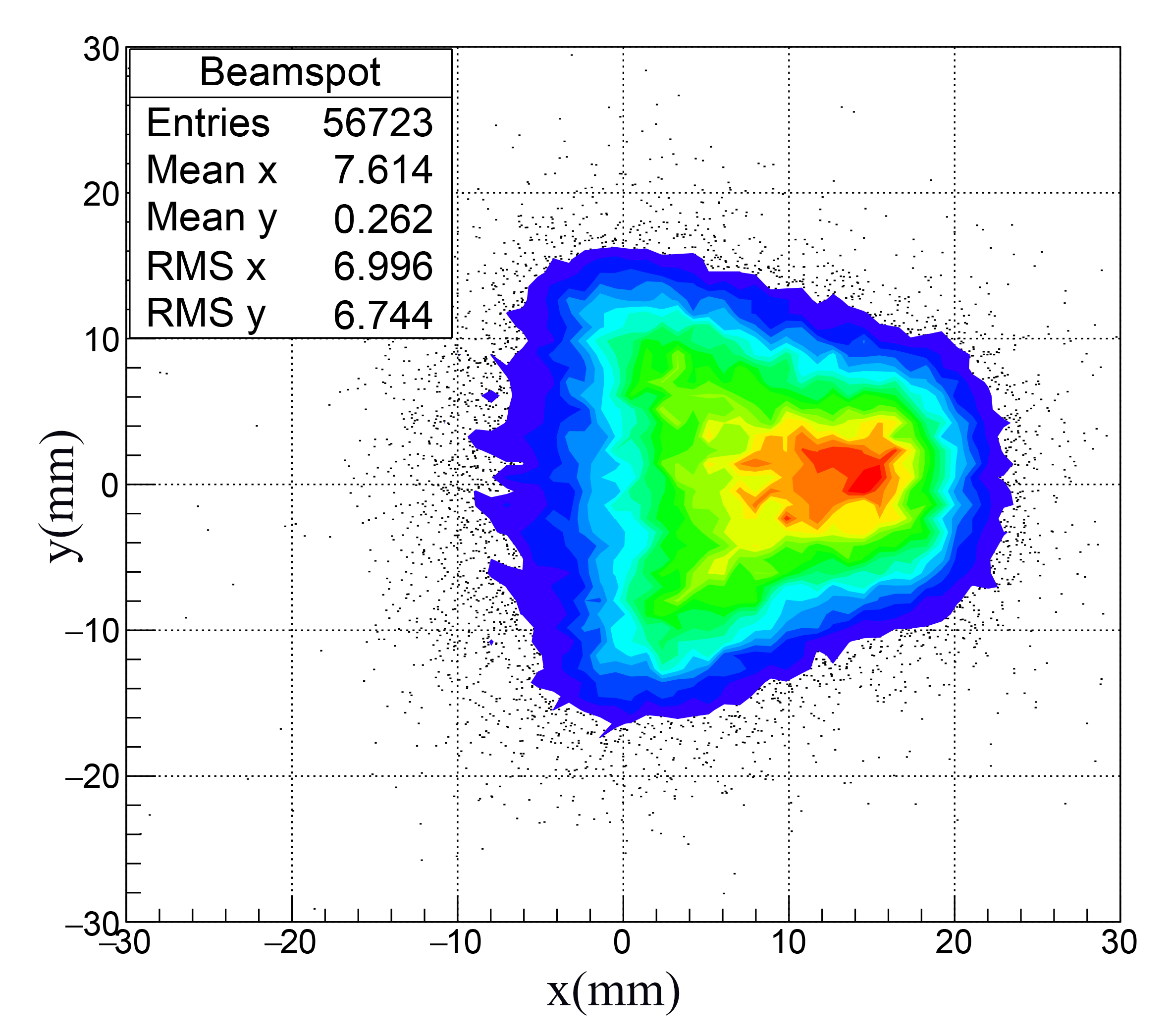}
\caption{Comparison of measured and simulated beam spots at the sample viewed from upstream with 
 RAT = RAB, and RAL-RAR = -0.64 kV (top), and 0.56 kV (bottom). Other parameters are : 
 V$_{mod}$=10.0 kV, L1= 6.0 kV, L2=7.0 kV, L3=8.0 kV, RAT=RAB=7.5 kV, $\theta_{spin}$= -10$^\circ$. 
 The initial number of muons at the moderator in the simulation is 10$^5$.}
\label{FIG7}
\end{figure*}
\begin{figure*}[ht]
\includegraphics[width=14 cm, height=8 cm]{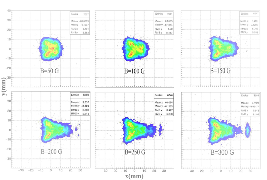}
\centering
\caption{Simulated beam spots at the sample position viewed from downstream with maximized number of muons in the center. 
The centering of the beam spot is obtained by using the RA potential differences given in Tab.~\ref{table1}. 
Vertical magnetic field (AEW magnet) varies from 50 Gauss to 300 Gauss with the same beam element settings as in 
Fig. \ref{FIG4} (V$_{mod}$=15 kV, RA=11.9 kV). Pictures from left to right, top to bottom are the beam spots
with different AEW magnetic fields (50 to 300 Gauss) at the sample with zero sample bias.}
\label{FIG8}
\end{figure*}

\begin{figure}[ht]
\centering
\includegraphics[width=7cm]{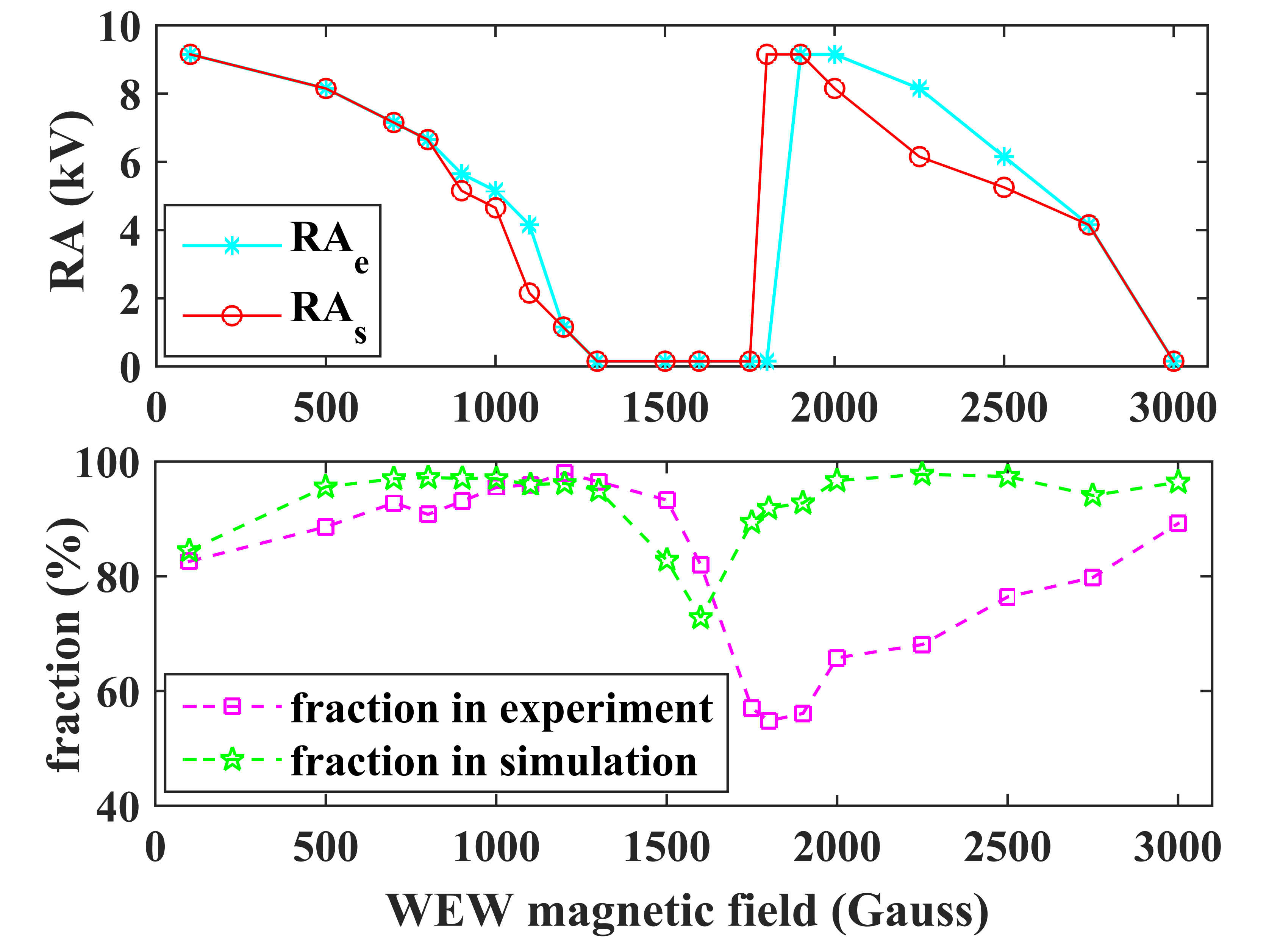}
\centering
\caption{Optimized RA potential values (top) and the corresponding fractions (bottom) of the beam stopping in the 
  central area 
  of 20$\times$20 mm$^2$ at the sample for different WEW magnetic fields for a moderator potential of 12 kV. 
  RA$_e$ and RA$_s$ are the best values in the experiments and simulations, respectively.}
\label{FIG9}
\end{figure}
\begin{figure}[ht]
\includegraphics[width=7cm]{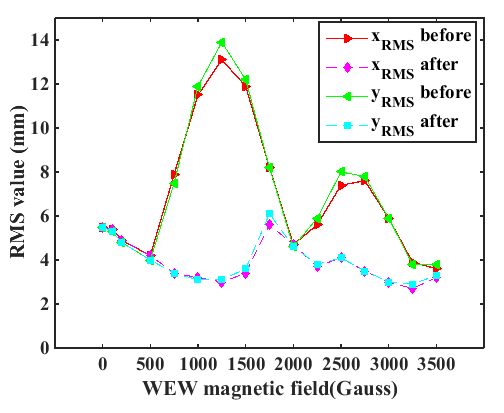}
\centering
\caption{$x_{RMS}$ and $y_{RMS}$ of the muon beam spot at the sample at different WEW magnetic fields 
  before and after tuning RA with V$_{mod}$=15 kV. Data labeled "before" and "after" are for RA=11.9 kV for 
  all magnetic fields before
  optimizing RA in the simulation and after setting the RA according to the values of Tab.~\ref{table4}, respectively.}
\label{FIG10}
\end{figure}
\begin{figure}[ht]
\includegraphics[width=7cm]{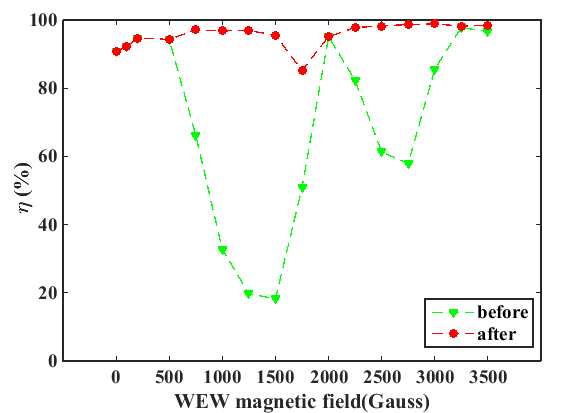}
\centering
\caption{The fraction $\eta$ of muons stopping in the central area of 20$\times$20 mm$^2$ at the sample position
  with different WEW magnetic fields parallel to the muon before and after optimizing RA with V$_{mod}$=15 kV. 
  Data labeled "before" and "after" are for RA=11.9 kV for all magnetic
  fields (before optimizing RA in the simulation) and after tuning RA according to Tab.~\ref{table4}, respectively. }
\label{FIG11}
\end{figure}

\subsection{\label{sec2-1} Beam spot with vertical magnetic field}
\subsubsection{\label{sec2-1-1}Effect of vertical magnetic field on beam spot}
In this section we investigate the effects of the vertical magnetic field (AEW magnet) and of an accelerating/decelerating 
electric field at the sample on the beam spot for different
fields and implantation energies. In addition, we simulate the steering and focusing effects of RA. 
This allows us to optimize the RA settings for the different magnetic field configurations, which can be used as
reference values for the experiment. Figure~\ref{FIG4} shows the simulated muon beam spot at the sample position
for a muon extraction voltage of V$_{mod}$ = 15 kV without external fields at the sample. 
The beam spot is well centered in this case, in agreement with the experiment (not shown). 
Figure~\ref{FIG5} shows beam spots without electric field at the sample, viewed from downstream for different 
AEW magnetic fields:
50, 100, 150, 200, 250 and 300 Gauss. The beam spot shifts almost linearly to the left with increasing magnetic field
while the shape of the beam spot is only marginally affected. The mean vertical position remains almost constant.
The mean horizontal position (meanX) for different electric fields and vertical magnetic fields are shown in
Fig. \ref{FIG6}. The absolute value of meanX increases nearly linearly with increasing magnetic field, and the beam
spot shift is larger in a decelerating electric field. In order to keep the beam spot in the center a potential
difference has to be applied between the RAL and RAR segments of RA. Its effect on the beam spot is studied in the
next subsection. 
\subsubsection{\label{sec2-1-2}Effect of beam steering by RA}
The effect of beam steering by applying a potential difference between opposite RA segments is
illustrated in Fig.~\ref{FIG7} for beam transport with V$_{mod}$ = 10 kV, where
we compare the experimental and simulated beam spots at the sample. 
In the experiment, the beam spot at the sample position is measured using a 44 mm diameter Roentdek 
DLD40 delay line detector with three resistance matched Hamamatsu MCPs in a Z configuration. 
The top two beam spots of Fig.~\ref{FIG7} are for RAL-RAR = -0.64 kV (RAL = 7.18 kV, RAR = 7.82 kV, RAT = RAB =7.50 kV), 
where the beam spot center shifts to the left side with meanX= -9.5 mm (in experiment) and -8.4 mm (in simulation). 
The results for RAL-RAR = 0.56 kV (RAL = 7.78 kV, RAR = 7.22 kV, RAT = RAB =7.50 kV) are shown in the bottom 
two figures. In this case the beam spot shifts to the right side with meanX= 8.8 mm (in experiment) and 7.6 mm 
(in simulation). The agreement between experiment and simulation is fairly good. 
The small deviations can be explained by the uncertainty in the applied potentials in the experiment.
The used high voltage power supplies have an absolute uncertainty between 10 V and 30 V, which results in
an overall uncertainty of about 50~V per power supply by taking into account additional uncertainties in the 
analog control of the power supplies. 
The beam spot is deformed to a trapezoidal shape, which is different from the effect of the magnetic field 
which doesn't cause a deformation. This deformation is caused by lens aberrations in the RA due to proximity
of the beam to the electrodes and the broken four fold symmetry of the electric field in case of a potential
difference applied between two opposite segments. 
In the simulation, the transmission from the moderator to the sample at 10 kV is about 55\%, compared to 
about 70\% at 15 kV, where we define the transmission as $\frac{N^{out}}{N^{in}}$ $\times$100\%, with $N^{out}$ the
number of muons in the sample plane, and $N^{in}$ the initial number of muons at the moderator. 
The reduction of the transmission is due to i) a longer time-of-flight at 10 kV, which increases the
fraction of muons decaying in flight by about 5\% at 10 kV compared to 15 kV, 
ii) a 10\% higher muonium formation (bound neutral state of a $\mu^+$ and an $e^-$) probability  in the 10-nm-thin 
carbon foil of the start detector, 
and iii) larger beam divergence at 10 kV after the start detector. 
In the experiment we observe a corresponding drop of event rate when changing the beam transport
 settings from 15 kV to 10 kV.
%
\begin{table}[ht]
\centering
\caption{Simulated optimum values of (RAL-RAR) (Unit: kV) to steer the beam spot to the center
at different electric potentials at the sample between -10 kV and 10 kV, and different AEW magnetic fields from 
0 to 300 Gauss, V$_{mod}$=15 kV. RAL+RAR=23.8 kV.}
\begin{tabular}{|c|c|c|c|c|c|}
\hline
  & -10 kV & -5 kV & 0 & 5 kV & 10 kV \\
\hline
50 G  & -0.36 & -0.38 & -0.40 & -0.50 & -0.6 \\
100 G & -0.68 & -0.70 & -0.76 & -0.82 & -0.9 \\
150 G & -1.00 & -1.10 & -1.20 & -1.20 & -1.2 \\
200 G & -1.25 & -1.50 & -1.70 & -1.60 & -1.6 \\
250 G & -1.60 & -1.70 & -1.85 & -2.00 & -2.0 \\
300 G & -2.00 & -2.10 & -2.20 & -2.30 & -2.4 \\
\hline
\end{tabular}
\label{table1}
\end{table}

\begin{table}[ht]
\centering
\caption{ The fraction $\eta$ of muons stopping in an area of 20$\times$20 mm$^2$ in the
center of the beam spot after steering by RA according to Tab. \ref{table1} for different
electric and AEW magnetic fields, V$_{mod}$=15 kV. (Unit:\%)}
\begin{tabular}{|c|c|c|c|c|c|}
\hline 
 & -10 kV &   -5 kV &  0 kV &  5 kV  & 10 kV\\  
 \hline 
0         & 93.7 & 93.2 & 91.9 & 90.4 & 87.6 \\ 
50 G  & 93.3 & 92.7 & 91.7 & 89.0 & 86.2 \\ 
100 G & 91.7 & 91.3 & 88.8 & 86.2 & 81.1 \\ 
150 G & 88.6 & 85.4 & 81.4 & 73.2 & 64.3 \\ 
200 G & 84.0 & 85.0 & 74.1 & 69.4 & 63.2 \\ 
250 G & 80.5 & 79.6 & 76.9 & 66.4 & 56.5 \\ 
300 G & 79.4 & 75.2 & 68.9 & 62.4 & 62.1 \\ 
\hline 
\end{tabular} 
\label{table2}
\end{table}

\subsubsection{\label{sec2-1-3}Centering of the  beam spot at the sample by RA steering}
%
\begin{table*}[ht]
\begin{center}
\caption{Optimized RA potential values for different WEW magnetic fields, $V_{mod}$=15 kV.}\label{table4}
\begin{tabular}{|c|c|c|c|c|c|c|c|c|c|c|c|c|c|c|c|}
	\hline
	B/Gauss & 100  & 200  & 500  & 750  & 1000 & 1250 & 1500 & 1750 & 2000 & 2250 & 2500 & 2750 & 3000 & 3250 & 3500 \\  \hline 
	RA/kV  & 11.9 & 11.9 & 11.9 & 10.0 & 7.6  & 3.8  &  0   &  0  & 11.2 & 10.5 & 7.2  & 4.2  & 0.0 & 0.0 & 11.0  \\ \hline
\end{tabular} 
\end{center}
\end{table*}

A potential difference between the RAL and RAR segments allows to counteract the action of the AEW magnetic 
field and to steer the beam spot back to the center at the sample position. 
For muons extracted with 15 kV at the moderator and whose beam spot shifts are shown in Fig. \ref{FIG5}, 
the optimum potential differences obtained by the simulation are shown in Tab.~\ref{table1} for different sample
potentials (i.e. implantation energies) and AEW magnetic fields. 
The other transport element parameters in the simulation are the same as those given in Fig. \ref{FIG4}. 

Figure~\ref{FIG8} shows the simulated beam spots after centering by using the RA parameters given in 
Tab.~\ref{table1}. The shape of the beam spot is deformed with increasing $|$RAL-RAR$|$ to a trapezoid 
instead of the initial square. When $|$RAL-RAR$|>2$~kV the beam spot splits into two parts, the larger
one keeping the previous trapezoidal shape, and the smaller one having an ellipsoidal shape. This splitting causes
the meanX value to deviate from zero, while the muon rate in the center of the sample plane is maximized. 

The fraction $\eta =  N^{out}_{[-10,10]mm}/N^{out}$ of muons stopping in an area of $x,y\in [-10, 10]$ mm 
in the center of the beam spot is shown in Tab.~\ref{table2} after RA steering. Before adjusting RA this
fraction is smaller than 10\% when a 300-Gauss magnetic field is 
applied, and it further reduces to less than 1\% when a positive bias of +5 kV and +10 kV is applied at the sample. 
After steering, most of the muons ($> 60$\%) are shifted to the 20$\times$20 mm$^2$ area in the center.

\subsection{\label{sec2-2} Beam spot with horizontal magnetic field}
In this section we investigate the effects of a horizontal magnetic field parallel to the beam axis 
on the beam spot for different fields and implantation energies. This magnetic field at the sample is
generated by the WEW magnet for LF- and TF-$\mu$SR experiments. The muon beam spot at the sample may be
influenced by the WEW magnetic field by field components perpendicular to the muon momentum. 
An experimental determination of the optimal RA settings is obtained by maximizing the fraction $\eta$ of muons 
landing within an area of 20$\times$20~mm$^2$ at the sample position. This can be done by measuring the
TF-$\mu$SR precession amplitude on a 20$\times$20~mm$^2$ gold foil glued onto a Ni coated large plate. 
Muons stopping in the gold foil maintain their polarization and hence contribute to the measured signal, 
while muons landing in the Ni backing depolarize almost immediately and do not contribute to the precession amplitude. 
Measurements of the precession amplitude as a function of RA voltage were performed using the positron detectors,
with V$_{mod}$ = 12 kV.
Figure~\ref{FIG9} shows good agreement between experimental (RA$_e$) and simulated (RA$_s$) optimum RA values for
WEW fields up to 1600 Gauss. At higher fields there are differences up to about 2500 Gauss, where the RA$_s$ values 
deviate by one to two kV from the experimental RA$_e$ optimum settings. At even higher fields RA$_e$ and RA$_s$ agree
again up to the maximum measured field of 3000 Gauss. We attribute the discrepancies to a slightly off-centered beam
in the experiment and possible differences between the calculated WEW field map used in the simulation and the
actual experimental one. The off-centered beam is caused by a slightly tilted muon moderator target and possible
misalignment of optical elements, which in turn introduces additional transverse velocity components in the beam, 
causing a shift and off-centered beam spot as a function of magnetic field. This affects the fraction of muons stopping
in the central area of 20$\times$20 mm$^2$.

The RMS values of simulated beam spots at different WEW magnetic field from 0-3500 Gauss at V$_{mod}$=15 kV are
displayed in Fig. \ref{FIG10}. For RA fixed at 11.9 kV the beam size significantly increases between 500 Gauss and
about 2000 Gauss due to the varying focusing power of the magnet. This can be corrected by lowering the RA potential
to reduce the focusing power of RA. As in the case of the V$_{mod}$~=~12~kV data the RA potential was tuned to maximize
the fraction $\eta$ of muons stopping in the central area of 20$\times$20 mm$^2$ of the sample plane.  
Table~\ref{table4} summarizes these optimized RA potential values to achieve the smallest beam spot size at the sample. 
One can see that the RA voltage can be kept constant up to 500 Gauss before it has to be lowered to compensate for 
the increasing focusing power of the magnet. At some magnetic fields (such as B=1500, 1700 Gauss) there is no need 
to use the RA to focus the beam (RA=0). Compared to the V$_{mod}$~=~12~kV data the RA has to be turned on again to
obtain the smallest beam spot at a field of about 2000~G, which is higher than the $\sim$~1750~G at V$_{mod}$~=~12~kV.
This shift is expected due to the $\sqrt{15~\rm{keV}/12~\rm{keV}}$ higher momentum of the muon beam: at higher beam
momentum one needs a correspondingly higher magnetic field to obtain the same beam transport properties of the
WEW magnetic field. 
Figure~\ref{FIG11} compares the fraction $\eta$ at the sample before and after
tuning the RA at different WEW magnetic fields. It is obvious that $\eta$ can be significantly increased 
by proper tuning of RA. The fractions $\eta$ at V$_{mod}$~=~15~kV are higher than at V$_{mod}$~=~12~kV because of the
smaller increase of transverse phase space when passing through the carbon foil of the start detector: the 
mean scattering angle due to multiple scattering is lower at higher beam energy.

\section{\label{sec4}Summary }
In this paper we presented the focusing and steering properties of a segmented conical electrostatic lens (RA)
which serves as a lens with large focusing power for the keV muon beam of the LE-$\mu^+$ facility at PSI. This 
beam optics element is essential for obtaining a small beam spot in the very limited space available in the
sample region of the LE-$\mu$SR setup. We presented a detailed Geant4 investigation of the beam transport
to optimize the experimental conditions for the present LEM setup, where we studied the beam transport onto
the sample plane in the presence of various magnetic and electric fields in the sample region. In some cases
the availability of experimental data allowed comparing the simulation with the experimental data. Good agreement
is found, which demonstrates that the optical properties of RA are well described in the simulation.
Using the simulation we optimized the electric potential settings of RA in the case of a vertical magnetic field
at the sample position. This field is transverse to the muon momentum and requires steering by RA to center
the beam spot. 
In the case of a magnetic field along the beam direction the increasing focusing power of the magnet has
to be compensated by a reduction of the RA focusing power. The simulation can be used to optimize RA for various
experimental conditions in the sample region without the need of running an experiment to test the beam properties
at the sample position for each case. This is important for the design and analysis of future 
LE-$\mu$SR experiments. A long term goal is the reduction of the beam spot size to allow the investigation of standard
5$\times$5~mm$^2$ samples. At present, the study of such samples is only possible by using a mosaic of at least
four pieces of this size. In many cases it is not possible to generate four or more identical samples, which
makes some experiments unfeasible. To achieve this long term goal the understanding and the reliability of the
simulation of the used optical beam elements is essential, especially the design of the last focusing element
where the present work provides important information.
Finally, we emphasize that Geant4 simulations are very powerful to describe and optimize experimental setups and
to help pushing experimental capabilities to the limit.

\section{\label{sec5}Acknowledgements}
Ran Xiao acknowledges a scholarship from the China Scholarship Council (CSC) and financial support from PSI for her stay at PSI.

\end{document}